\documentclass{article}

\usepackage{microtype}
\usepackage{graphicx}
\usepackage{subcaption}
\usepackage{booktabs} 

\usepackage{hyperref}

\usepackage[accepted]{icml2026}

\usepackage{amsmath}
\usepackage{amssymb}
\usepackage{mathtools}
\usepackage{amsthm}

\usepackage{natbib}
\usepackage{multirow}
\usepackage{placeins}
\usepackage{enumitem}
\usepackage{listings}
\usepackage[bottom]{footmisc}
\usepackage[most]{tcolorbox}
\tcbuselibrary{breakable, skins}

\lstdefinestyle{prompt}{
    basicstyle=\ttfamily\scriptsize,    
    breaklines=true,
    breakatwhitespace=false,
    columns=flexible,
    showstringspaces=false,
    keepspaces=true,
    lineskip=0pt,
    aboveskip=2pt,
    belowskip=2pt,
}
\usepackage[capitalize,noabbrev]{cleveref}

\theoremstyle{plain}

\theoremstyle{definition}

\theoremstyle{remark}

\usepackage[textsize=tiny]{todonotes}

\icmltitlerunning{MedJudgeRAG}

\begin{document}

\twocolumn[
  \icmltitle{MedJudgeRAG: Option-Wise Evidence Judgment with \\
  Dynamic Knowledge Graphs for Medical MCQA}



  \icmlsetsymbol{equal}{*}

  \begin{icmlauthorlist}
    \icmlauthor{Seongwon Seo}{no1}
    \icmlauthor{Seung Hwan Cho}{no1}
    \icmlauthor{Young-Min Kim}{no1,no2}
  \end{icmlauthorlist}

  \icmlaffiliation{no1}{Department of Industrial Data Engineering, Hanyang University, Seoul, South Korea}
  \icmlaffiliation{no2}{School of Interdisciplinary Industrial Studies, Hanyang University, Seoul, South Korea}

  \icmlcorrespondingauthor{Young-Min Kim}{yngmnkim@hanyang.ac.kr}

  \icmlkeywords{Machine Learning, ICML}

  \vskip 0.3in
]



\printAffiliationsAndNotice{}  

\begin{abstract}
  In medical multiple-choice question answering (MCQA), Retrieval-Augmented Generation (RAG) can supplement the domain knowledge of language models (LMs). However, since vanilla RAG indiscriminately utilizes retrieved documents, it can degrade LM performance. To address this, we propose MedJudgeRAG. Our framework represents retrieved documents as a dynamic knowledge graph (KG) composed of entities and relations. For each option, the model judges an evidence verdict from the retrieved documents and the KG. Based on the verdict combination, the model determines a knowledge utilization strategy to reason toward the final answer. These capabilities are trained via supervised fine-tuning using structured reasoning traces generated by a teacher LM. The training employs a weighted cross-entropy loss that differentially weights the KG and reasoning segments. Experiments on two medical MCQA benchmarks demonstrate that MedJudgeRAG consistently outperforms both vanilla RAG and parametric baselines. Furthermore, ablation analysis reveals that the dynamic KG is more effective as graph-conditioned supervision at training time than as an explicit output at inference time. Our code is available at \url{https://github.com/hyu-amllab/medjudgerag}, and the generated reasoning traces are released at \url{https://huggingface.co/datasets/youarethewon/medjudgerag}.
\end{abstract}

\section{Introduction}
\label{introduction:introduction}
Medical question answering (QA) has been studied across a wide range of evaluation settings, including medical examinations and biomedical research \cite{singhal2025toward}. Recent benchmarks that demand expert-level medical knowledge and advanced reasoning have further expanded medical QA toward assessing domain knowledge-grounded reasoning capabilities \cite{zuo2025medxpertqa}. Medical multiple-choice question answering (MCQA) benchmarks likewise include clinical vignette-based questions, with answer options designed to require fine-grained medical understanding \cite{jin2021disease, pal2022medmcqa}. In such settings, simple fact recall is insufficient to distinguish the correct answer. Instead, acquiring additional domain knowledge or providing useful external evidence for reasoning becomes essential.

Pre-trained language models (LMs) rely primarily on parametric knowledge acquired during training, which limits their performance on tasks that require domain-specific reasoning. To address this limitation, Retrieval-Augmented Generation (RAG) has been proposed \cite{lewis2020retrieval, guu2020retrieval}. RAG retrieves relevant documents from an external corpus and provides them as input to the LM. This supplements the model's limited domain knowledge and has demonstrated potential for improving performance on domain-specific QA tasks. However, when applying RAG to medical MCQA, vanilla RAG, which simply prepends retrieved documents to the input, still exhibits notable limitations. When the evidence for the correct answer is scattered across multiple documents, the model struggles to structurally connect these fragments. Moreover, it lacks a mechanism to systematically judge whether the retrieved information constitutes sufficient evidence for each option. Finally, it lacks an explicit intermediate decision process to bridge retrieved evidence and final answer selection. As a result, these limitations can cause the reasoning accuracy of retrieval-augmented LMs to fall below that of LMs relying solely on parametric knowledge. To overcome these challenges, the LM must be able to structure scattered evidence, explicitly judge whether sufficient evidence exists for each option, and determine how to utilize retrieved information based on these judgments.

In this work, we introduce MedJudgeRAG, a novel framework in which the LM learns to leverage a dynamic knowledge graph for option-wise evidence judgment to reason toward the final answer. MedJudgeRAG trains the LM through supervised fine-tuning (SFT) to acquire two capabilities. In Step 1, the LM learns to extract key entities and relations from retrieved documents that are discriminative for the given question and options. This structures the external knowledge into a dynamic KG. In Step 2, the LM learns to identify option-wise evidence based on the retrieved documents and the KG. It then judges whether this evidence supports each option. Based on these judgments, the model determines how to leverage external knowledge for answer reasoning . Specifically, when sufficient evidence supports a particular option, the model reasons directly from that evidence. When evidence only contradicts certain options, the model applies elimination to narrow down candidates. When the evidence is insufficient for all options, the model reverts to parametric knowledge. To train these capabilities, we leverage structured reasoning traces generated by a teacher LM as SFT data. The fine-tuned model takes a question, options, and retrieved documents as input. It then performs dynamic KG construction, option-wise evidence judgment, knowledge utilization strategy selection, and final answer prediction in an end-to-end manner.

We evaluate MedJudgeRAG on the MedQA and MedMCQA benchmarks \cite{jin2021disease, pal2022medmcqa}, with parametric inference and vanilla RAG as baselines under the same backbone and identical retrieval results. A notable finding is that MedJudgeRAG mitigates the known problem of performance degradation caused by indiscriminately utilizing retrieved documents. This improvement does not stem from learning additional domain knowledge. Instead, the model learns dynamic knowledge construction and a reasoning strategy tailored to multiple-choice questions. Furthermore, through ablation analysis, we find that the dynamic KG functions primarily as structural training-time supervision rather than a required inference-time output. Accordingly, we explore both explicit and implicit KG decoding and analyze the trade-off between them. Our main contributions are as follows:
\begin{itemize}
	\item We propose MedJudgeRAG, a framework that structures retrieved documents into a dynamic KG and performs MCQA-focused evidence reasoning. Experiments on two medical MCQA benchmarks demonstrate that MedJudgeRAG overcomes the limitations of vanilla RAG under the same retrieval conditions.
	\item Through ablation analysis, we reveal that the dynamic KG contributes primarily as graph-conditioned supervision at training time rather than as an explicit output at inference time. We further analyze the trade-off between explicit and implicit KG generation, showing that each mode has complementary strengths.
\end{itemize}

\section{Related Work}

\subsection{Retrieval-Augmented Generation}

Retrieval-Augmented Generation (RAG) can supplement the domain knowledge deficiency of generator models by augmenting the prompt of pre-trained models with retrieved documents \cite{lewis2020retrieval, guu2020retrieval}. Recently, advanced RAG frameworks have been proposed. These frameworks enable the generator LM to discriminately accept retrieved documents. \citet{jiang2023active} proposed a framework that determines when and what to retrieve during generation through an active retrieval strategy that triggers retrieval only when the LM generates low-probability tokens. \citet{lin2023ra} introduced a two-stage fine-tuning approach that enables the retrieval-augmented LM to better utilize retrieved information while aligning the retriever with the LM's preferences to yield more contextually relevant results. \citet{asai2023self} employed special tokens to determine whether external knowledge retrieval is necessary and to self-criticize the LM's outputs, thereby selecting the optimal generation. \citet{wei2024instructrag} demonstrated that LMs can explicitly remove noise from retrieved documents by generating denoising rationales. Our work discriminately incorporates retrieved documents by leveraging a dynamic KG to guide option-wise evidence judgment.

\subsection{Knowledge Graphs in Retrieval-Augmented Generation}

A knowledge graph (KG) is a structured representation of facts composed of entities, relationships, and semantic descriptions \cite{ji2021survey}. Recently, KGs have been adopted in RAG as a means of structuring external knowledge. Related work can be categorized into approaches that structure external knowledge as a graph before inference and those that do so at inference time. Methods in the former category convert a corpus or multiple documents into a pre-structured knowledge source. \citet{wang2024knowledge} construct a KG from passages using nodes that represent passages across multiple documents and edges that represent connections between nodes based on similarity and lexical overlap. GraphRAG leverages LLMs to construct a KG from a large text corpus \cite{edge2024local}. HippoRAG transforms a corpus into a schemaless knowledge graph \cite{gutierrez2024hipporag}. Methods in the latter category convert retrieved evidence into a KG tailored to the current question. REANO proposes a module that generates a KG composed of a set of knowledge triples from retrieved passages \cite{fang2024reano}. RAS constructs a question-specific KG through iterative retrieval \cite{jiang2025ras}. In the medical domain, MedGraphRAG proposes a knowledge graph construction method based on triple graph construction along with an efficient retrieval method called U-Retrieval \cite{wu2025medical}. However, since MedGraphRAG pre-constructs the graph and retrieves from this structured knowledge, its KG construction and utilization approach differs from that of our work. TAdaRAG constructs an on-the-fly KG using prompt templates tailored to the question domain, but represents only entities explicitly \cite{zhang2026tadarag}. In contrast, our KG explicitly represents both entities and relationships, and is leveraged as training data.

\subsection{LLMs for Medical Question Answering}

Medical question answering often requires selecting the correct answer from multiple candidates, typically in a multiple-choice or selection-based format \cite{jin2019pubmedqa, hendrycks2021mmlu}. To improve LLM performance on medical QA, prior work has primarily pursued two directions, namely training additional medical domain knowledge or leveraging external knowledge through RAG. A representative approach for the former is adapting general-purpose LLMs to medical corpora or clinical data \cite{labrak2024biomistral, chen2023meditron, kim2025small, christophe2024med42}. For the latter direction, $\text{RAG}^2$ focuses on the retriever side of medical QA and introduces rationale-guided filtering to mitigate noisy retrieval and poorly targeted queries \cite{sohn2025rationale}. Self-BioRAG is a framework designed for reliable biomedical text processing that generates explanations, retrieves domain-specific documents, and self-reflects on generated responses \cite{jeong2024improving}. However, these medical RAG methods presuppose adaptive retrieval or modifications to the retrieval pipeline. Our work aims to better utilize retrieved external knowledge within a fixed retrieval setting by explicitly training the model on the process of judging an evidence verdict for each option.

\section{Method}
\label{method:method}
\begin{figure*}[t]
\centering
\includegraphics[width=\textwidth]{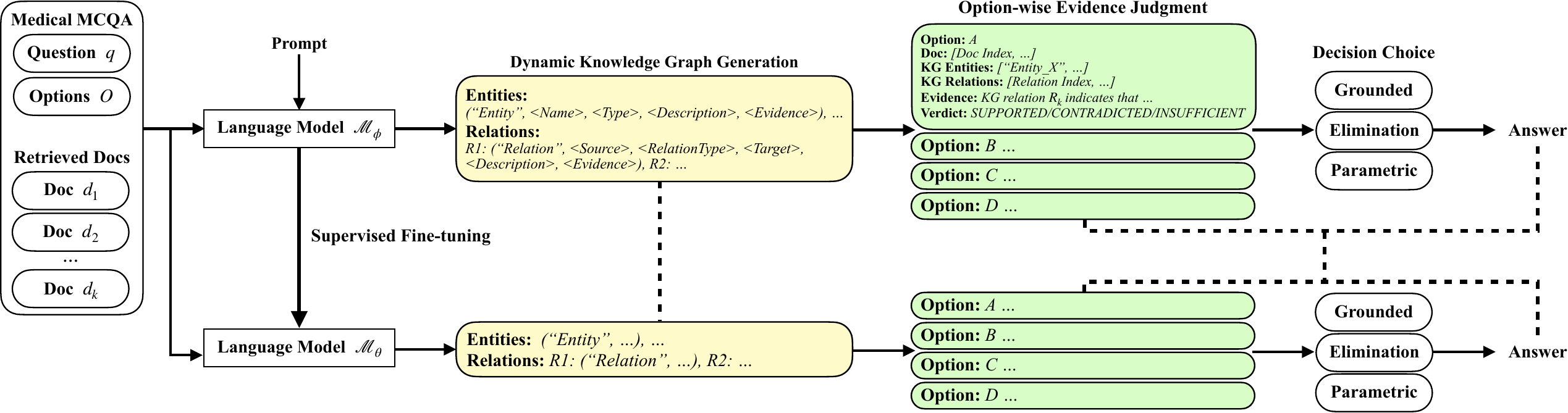}
\caption{Overview of the MedJudgeRAG framework. 
\textbf{Top}: the teacher LM $\mathcal{M}_\phi$ (GPT-5.1) generates structured reasoning traces consisting of a dynamic knowledge graph, option-wise evidence judgments, and a decision. \textbf{Bottom}: the student LM $\mathcal{M}_\theta$ (Mistral/Llama) is trained via supervised fine-tuning on the teacher's outputs and produces the same structured output at inference time. Dashed lines indicate that the teacher's outputs serve as SFT training targets.}
\label{fig:pipeline}
\end{figure*}

We propose MedJudgeRAG for medical MCQA. MedJudgeRAG represents retrieved documents as a dynamic knowledge graph. The model judges an evidence verdict for each option and reasons toward the correct answer. Figure~\ref{fig:pipeline} illustrates the overall pipeline of MedJudgeRAG. Our end-to-end training first enables the LM $\mathcal{M}$ to generate a dynamic knowledge graph from the question, options, and retrieved documents. The model then generates an $\texttt{<ANALYSIS>}$ delimiter to initiate option-wise evidence analysis. After judging whether each option is supported by the external knowledge or the KG, the model determines its reasoning strategy based on the judgment results.

\subsection{Problem Formulation}
\label{method:problem_formulation}
LM $\mathcal{M}_\theta$ has access to an annotated dataset for the Multiple Choice Question Answering task $\mathcal{T} = \{ \langle q, O, a \rangle \}$, where $q$ is a question, $O = \{ o_1, \dots, o_n \}$ is an option set, and $a$ is the answer. In the standard RAG setting, the model additionally has access to an external knowledge base via a retriever $\mathcal{R}$. Given a question $q$, the retriever $\mathcal{R}$ returns a set of relevant documents $D = \{d_1, \dots, d_k\}$ from the external knowledge base. The model selects the most appropriate $o_i$ from the option set $O$ for the given question $q$ based on $D$ and its own parametric knowledge, denoted as $p_\theta(a | q, O, D)$.

MedJudgeRAG trains $\mathcal{M}_\theta$ to sequentially generate a single generation sequence $y = [g; r]$ from $x = (q, O, D)$. Here, $g$ is a dynamic knowledge graph extracted from $D$ based on $q$ and $O$, consisting of entity tuples and relation tuples. The MCQA-focused reasoning sequence $r$ denotes the sequential generation of $[j; c; a]$. $j$ denotes the evidence judgment, which assigns a verdict $v \in \{\text{supported}, \text{contradicted}, \text{insufficient}\}$ for each option. The decision $c \in \{\text{grounded}, \text{elimination}, \text{parametric}\}$, which determines the knowledge utilization strategy, is conditioned on $j$. The final answer $a$ is then produced according to $c$.

\subsection{Dynamic Knowledge Graph Construction}
\label{method:dynamic_knowledge_graph_construction}
Retrieved documents can be noisy, and evidence for the correct answer is often scattered across multiple documents. When such documents are directly fed to the LM, the model may be distracted by irrelevant information or fail to connect dispersed evidence. To address this, we employ a pre-designed extraction template to extract only the information relevant to the question and options from the retrieved documents and structure it as a dynamic knowledge graph.

\textbf{KG Schema and Format.} The KG in MedJudgeRAG consists of two types of elements, entities and relations, each represented as a structured tuple:

\noindent
{\small
\textbullet\ Entity tuple: (\texttt{"Entity"}, Name, Type, Description, Evidence)\\
\textbullet\ Relation tuple: (\texttt{"Relation"}, Source, RelationType, Target, Description, Evidence)
}

Entity types follow the 15 Unified Medical Language System (UMLS)\footnote{\url{https://www.nlm.nih.gov/research/umls}} Semantic Groups, while relation types are adapted from UMLS relations with task-specific adjustments. These schema constraints serve to reduce redundancy and improve entity relevance \cite{zhang2026tadarag}. The Evidence field lists the source document indices for each element (e.g., [1] or [1, 2]).

\textbf{KG Data Generation.} To construct training data, we provide the teacher LM $\mathcal{M}_\phi$, GPT-5.1 \cite{openai2025gpt51}, with $(q, O, D)$. We then prompt it to generate a KG following the KG Generation Principles specified in the instruction, along with a one-shot example.

\begin{itemize}[itemsep=0pt, parsep=0pt, topsep=0pt]
	\item \textbf{Discriminative extraction.} Rather than extracting all medical concepts appearing in the retrieved documents, only information that contributes to distinguishing among the four options is selectively extracted.
	\item \textbf{Document-grounded.} All entity names and descriptions must be grounded in the retrieved documents. This prevents the extraction of hallucinated entities that rely on the model's own parametric knowledge. Relations are extracted only when explicitly supported by the documents.
	\item \textbf{Evidence-traced.} All entities and relations specify their source document indices. This enables tracing the basis of verdicts in the subsequent judgment stage, strengthening evidence-grounded reasoning.
	\item \textbf{Empty graph allowed.} When the retrieved documents are irrelevant to the question or contain only information unhelpful for answer determination, the model may output an empty KG. This prevents amplifying noise by forcing extraction from irrelevant documents.
\end{itemize}

To ensure that the KG is generated under the same conditions as $\mathcal{M}_\theta$ at inference time, the gold answer is not included in the teacher prompt. The generated KG undergoes the following schema-aware post-processing steps before being adopted as training data: (1) normalizing entity types to the pre-defined 15 categories, (2) filtering invalid relation types, and (3) removing relations whose source or target entities do not exist among the extracted entities.

\subsection{Option-wise Evidence Judgment}
\label{method:option-wise_evidence_judgment}
To address the limitations of vanilla RAG discussed in Section~\ref{introduction:introduction}, we train the model with an explicit generation structure that judges an evidence verdict for each option, determines a reasoning strategy based on the verdict combination, and then selects the answer.

\textbf{Option-wise Verdict Generation.} The model is trained to first output an $\texttt{<ANALYSIS>}$ delimiter after KG generation. $\texttt{<ANALYSIS>}$ delimiter signals the beginning of MCQA-focused reasoning generation and is included in $r$. Following this delimiter, the model sequentially generates a verdict block for each option $o_i \in O$. Each verdict block describes the supporting evidence for the option by referencing relevant document indices, entity names, and relation indices. A verdict is then selected based on the written evidence. This evidence-first design suppresses unsupported verdicts by requiring the model to articulate evidence before committing to a judgment. The verdict consists of three labels: (1) \textbf{SUPPORTED}: The documents or KG provide direct evidence supporting the option as the correct answer. (2) \textbf{CONTRADICTED}: The documents or KG explicitly indicate that the option is incorrect. (3) \textbf{INSUFFICIENT}: The documents or KG alone do not provide sufficient basis to determine whether the option is correct or incorrect.

\textbf{Three-Way Decision Choice.} Once the option-wise verdicts are determined, the model selects one of three knowledge utilization strategies based on the combination of verdicts. This decision explicitly branches how the retrieved information is utilized. The decision consists of three labels:

\begin{itemize}[itemsep=0pt, parsep=0pt, topsep=0pt]
	\item \textbf{Grounded}: When a SUPPORTED verdict exists. The model selects the SUPPORTED option as the answer based on the retrieved evidence.
	\item \textbf{Elimination}: When no SUPPORTED verdict exists but at least one CONTRADICTED verdict is present. The model eliminates the CONTRADICTED options as incorrect and reasons over the remaining options with the aid of parametric knowledge.
	\item \textbf{Parametric}: When all options receive an INSUFFICIENT verdict. The retrieved information is deemed insufficient for answer determination. The model selects the answer based on its own parametric knowledge.
\end{itemize}

\subsection{Training}
\label{method:training}
\textbf{Teacher Data Generation.} The training data consists of complete sequences $s^* = [g^*; r^*]$ generated by $\mathcal{M}_\phi$ via the OpenAI Batch API with greedy decoding (\texttt{temperature}=0), producing one trace per example without resampling. $\mathcal{M}_\phi$ receives only $(q, O, D)$ as input; the gold answer is excluded from the prompt and used only during post-hoc validation. Following instructions that specify the KG generation principles (Section~\ref{method:dynamic_knowledge_graph_construction}) and the MCQA-focused reasoning output structure (Section~\ref{method:option-wise_evidence_judgment}), $\mathcal{M}_\phi$ sequentially generates a KG, option-wise evidence judgment, decision, and answer. We draw 2,750 training samples from each of the MedQA and MedMCQA training splits with a fixed random seed. We evaluate on the held-out MedQA test split and the MedMCQA validation split. The MedMCQA validation split, disjoint from its training split, serves as the test set following the MedRAG convention since the official test labels are withheld. Each generated trace undergoes two stages of validation. Format validation checks structural constraints such as tag completeness, presence of required fields, and consistency of verdict-decision combinations. Semantic 
validation checks rule-based consistency between evidence and verdicts, adherence to decision rules, internal consistency between the final answer and the teacher's analysis, and agreement with the gold label. We further remove samples containing hint-leakage expressions that directly reveal the answer (e.g., ``the correct answer is'', ``obviously option A''). Of the 5,500 generated traces, 3,562 (64.8\%) passed validation. After removing samples exceeding 8,192 tokens under the Mistral-7B-Instruct-v0.3 \cite{jiang2023mistral7b} and Meta-Llama-3-8B-Instruct \cite{grattafiori2024llama} tokenizers, 3,479 (3,148 train / 331 val) and 3,559 (3,222 train / 337 val) samples were adopted for each backbone, respectively.

\textbf{Weighted Loss Fine-tuning.} MedJudgeRAG is trained to generate the generation sequence $y = [g; r]$ for all instruction-target pairs $(x, y)$ in the training set $\mathcal{D}_{\text{train}}$. However, since $g$ targets accurate KG extraction while $r$ targets the reasoning sequence, the two subsequences have different learning objectives. We therefore allocate more training focus to the reasoning segment and reduce the per-token gradient for the KG segment, which serves as an intermediate representation for reasoning, through loss weight adjustment. To this end, we employ a weighted cross-entropy loss that differentially weights tokens based on their position within the sequence. Let $\mathcal{T}_g, \mathcal{T}_r \subseteq \{1, \dots, |y|\}$ denote the index sets of token positions belonging to the KG segment and the reasoning segment, respectively ($\mathcal{T}_g \cap \mathcal{T}_r = \emptyset$). Their respective counts of valid tokens are denoted by $|\mathcal{T}_g|$ and $|\mathcal{T}_r|$. Prompt and padding tokens are excluded from the loss. The boundary between the two segments is determined by $\texttt{<ANALYSIS>}$ delimiter. The training objective is defined as follows:
\begin{equation}
\begin{aligned}
    \mathcal{L}(\theta) = -\mathbb{E}_{(x,y)\sim\mathcal{D}} \bigg[ \frac{1}{Z} \bigg( &\lambda_g \sum_{t \in \mathcal{T}_g} \log p_\theta(y_t \mid y_{<t}, x) \\
    &+ \sum_{t \in \mathcal{T}_r} \log p_\theta(y_t \mid y_{<t}, x) \bigg) \bigg]
\end{aligned}
\end{equation}
$\text{where } Z = \lambda_g|\mathcal{T}_g| + |\mathcal{T}_r|.$

The hyperparameter $\lambda_g \in [0, 1]$ controls the loss weight for the KG segment, while the loss weight for the reasoning segment is fixed at 1. Normalizing by the weighted sum $\lambda_g|\mathcal{T}_g| + |\mathcal{T}_r|$ fixes the per-token contribution ratio of the two segments (KG : reasoning) at $\lambda_g : 1$, independent of sample length.

\section{Experiments}

\subsection{Experimental Setup}
\label{experiments:experimental_setup}
\textbf{Datasets \& Metrics.} We evaluate on two medical MCQA benchmarks: the test set of MedQA (1,273 questions) and the validation set of MedMCQA (4,183 questions). Performance is measured by accuracy, where the predicted answer is extracted from the model's generation via regular expression parsing.

\textbf{Baselines.} All comparisons are conducted under the same backbone and the same retrieval results, varying only how the retrieved information is utilized. Parametric provides only the question and options without retrieved documents. Vanilla RAG appends the top-5 retrieved documents to the prompt via simple concatenation. Both Parametric and Vanilla RAG follow the chain-of-thought prompt from \citet{xiong-etal-2024-benchmarking}. The maximum generated output length is set to 1,024 tokens for Parametric and 8,192 tokens for Vanilla RAG to accommodate our experimental setting.

\textbf{Implementation Details.} To ensure a fair comparison, vanilla RAG and MedJudgeRAG use identical retrieval results. The retrieval corpus consists of PubMed\footnote{\url{https://pubmed.ncbi.nlm.nih.gov/}} for biomedical abstracts and medical textbooks for domain-specific knowledge \cite{jin2021disease}. The retriever is Contriever \cite{izacard2021unsupervised}, retrieving the top-5 documents per question. At inference time, $\mathcal{M}_\theta$ receives $(q, O, D)$ as input and operates in one of two decoding modes. In \textbf{Explicit decoding}, the model outputs KG, option-wise evidence judgment, decision, and answer in a single generation step. In \textbf{Implicit decoding}, KG generation is skipped and only option-wise evidence judgment, decision, and answer are generated. The comparison between Explicit and Implicit decoding is presented in Section ~\ref{ablation_study:ablation_study} to isolate the effect of KG generation. The maximum generated output length is 8,192 tokens, and we use vLLM to accelerate inference \cite{kwon2023efficient}. We adopt Mistral-7B-Instruct-v0.3 (hereafter Mistral) and Meta-Llama-3-8B-Instruct (hereafter Llama) as backbones. Training is performed via QLoRA-based SFT with ZeRO-2 \cite{dettmers2023qlora}. The maximum input sequence length is 8,192 tokens, per-GPU batch size is 1, and gradient accumulation is set to 16 steps. We train for 3 epochs with bfloat16 mixed precision, an initial learning rate of 1e-4, AdamW optimizer, and cosine scheduler. We experiment with KG loss weight $\lambda_g \in \{0.0, 0.1, 0.3, 0.5, 1.0\}$ and select the best-performing $\lambda_g$ based on average accuracy. Training takes 10--12 hours on 2 NVIDIA GeForce RTX 3090 (24\,GB) GPUs.

\begin{table}[t]
\centering
\small
\setlength{\tabcolsep}{5pt}
\caption{Accuracy on MedQA and MedMCQA. Mistral denotes Mistral-7B-Instruct-v0.3; Llama denotes Meta-Llama-3-8B-Instruct. All \textit{Ours} results use $\lambda_g=0.0$. Explicit decoding generates the KG before reasoning; Implicit decoding skips KG generation. \underline{Underline} denotes the best result per backbone.}
\label{tab:main_results}
\begin{tabular}{llccc}
\toprule
\textbf{Backbone} & \textbf{Method} & \textbf{MedQA} & \textbf{MedMCQA} & \textbf{Avg} \\
\midrule
\multirow{4}{*}{Mistral} 
  & Parametric        & 52.55          & 46.26          & 49.41 \\
  & Vanilla RAG       & 50.59          & 41.57          & 46.08 \\
  \cmidrule(l){2-5}
  & Ours (Explicit)   & 59.54          & \underline{50.37} & 54.96 \\
  & Ours (Implicit)   & \underline{60.88} & 50.16          & \underline{55.52} \\
\midrule
\multirow{4}{*}{Llama} 
  & Parametric        & 60.64          & 54.86          & 57.75 \\
  & Vanilla RAG       & 49.88          & 45.61          & 47.75 \\
  \cmidrule(l){2-5}
  & Ours (Explicit)   & 63.94          & 55.39          & 59.67 \\
  & Ours (Implicit)   & \underline{66.93} & \underline{57.33} & \underline{62.13} \\
\bottomrule
\end{tabular}
\end{table}

\subsection{Results}
\label{experiments:results}
\textbf{Main Results.} Table~\ref{tab:main_results} reports the accuracy of Parametric, Vanilla RAG, and MedJudgeRAG (Explicit/Implicit) for both backbones. The primary comparison is Vanilla RAG vs. Ours, measuring the effectiveness of MedJudgeRAG over indiscriminately utilizing retrieved information. Additionally, we compare Ours (Explicit) and Ours (Implicit) to analyze how explicitly generating the KG affects the reasoning chain. As shown in Table~\ref{tab:main_results}, Vanilla RAG degrades accuracy relative to Parametric on both backbones: by 2.0 percentage points (pp) and 4.7pp on MedQA and MedMCQA for Mistral, and by 10.8pp and 9.3pp for Llama. This confirms that simply prepending retrieved documents can degrade performance for these 7-8B-scale backbones. However, MedJudgeRAG overcomes this limitation using the same retrieved documents, achieving improvements over Vanilla RAG of up to +10.3pp (MedQA) and +8.8pp (MedMCQA) for Mistral, and +17.1pp (MedQA) and +11.7pp (MedMCQA) for Llama. Furthermore, MedJudgeRAG consistently improves over Parametric, demonstrating that it fulfills the original goal of RAG, enhancing LM performance by incorporating retrieved documents.

\textbf{Explicit vs. Implicit Decoding.} We compare Explicit decoding, which outputs the KG before reasoning, with Implicit decoding, which skips KG generation. For Mistral, Implicit decoding achieves higher accuracy on MedQA while Explicit decoding slightly outperforms on MedMCQA, showing that the advantage varies by benchmark. For Llama, Implicit decoding consistently outperforms Explicit decoding on both benchmarks. We analyze the trade-off between Explicit and Implicit decoding in detail in Section~\ref{ablation_study:ablation_study}.

\section{Ablation Study}
\label{ablation_study:ablation_study}
This section examines performance changes when KG generation is directly optimized ($\lambda_g > 0$) versus not ($\lambda_g = 0$). We also compare \textbf{Explicit} and \textbf{Implicit} decoding. Through these analyses, we show that the KG serves more effectively as graph-conditioned supervision at training time than as an explicit output at inference time. Furthermore, while KG learning intensity can be controlled via $\lambda_g$, improving KG surface-form quality does not directly translate to improved answer accuracy.

\begin{table}[t]
\centering
\footnotesize
\setlength{\tabcolsep}{2pt}
\caption{Ablation on KG loss weight per benchmark. \textit{w/o KG} is trained without the KG segment (independent of $\lambda_g$). \textit{Exp.}/\textit{Imp.} denote Explicit/Implicit decoding. \textbf{Bold}: best within each decoding for each (backbone, benchmark). 
\underline{Underline}: best per (backbone, benchmark).}
\label{tab:ablation_kgw}
\begin{tabular}{llcccccc}
\toprule
\textbf{Bench.} & \textbf{Method} & \multicolumn{5}{c}{\textbf{KG loss weight} ($\boldsymbol{\lambda_g}$)} & \multirow{2}{*}{\textbf{w/o KG}} \\
\cmidrule(l){3-7}
& & \textbf{0.0} & \textbf{0.1} & \textbf{0.3} & \textbf{0.5} & \textbf{1.0} & \\
\midrule
\multicolumn{8}{l}{\textit{Mistral}} \\
\multirow{2}{*}{MedQA}
  & Exp. & \textbf{59.54} & 53.73 & 54.05 & 57.34 & 55.15 & \multirow{2}{*}{55.15} \\
  & Imp. & 60.88 & \underline{\textbf{61.19}} & 60.17 & 59.23 & 60.49 & \\
\cmidrule(l){1-8}
\multirow{2}{*}{MedMCQA}
  & Exp. & \underline{\textbf{50.37}} & 42.96 & 44.68 & 46.52 & 46.12 & \multirow{2}{*}{47.57} \\
  & Imp. & \textbf{50.16} & 49.39 & 48.96 & 49.06 & 48.82 & \\
\midrule
\multicolumn{8}{l}{\textit{Llama}} \\
\multirow{2}{*}{MedQA}
  & Exp. & \textbf{63.94} & 57.89 & 58.05 & 60.49 & 59.94 & \multirow{2}{*}{62.29} \\
  & Imp. & \underline{\textbf{66.93}} & 65.59 & 64.65 & 63.32 & 63.16 & \\
\cmidrule(l){1-8}
\multirow{2}{*}{MedMCQA}
  & Exp. & \textbf{55.39} & 53.14 & 53.77 & 53.65 & 52.09 & \multirow{2}{*}{55.87} \\
  & Imp. & \underline{\textbf{57.33}} & 56.13 & 56.01 & 55.77 & 54.82 & \\
\bottomrule
\end{tabular}
\end{table}

\begin{table*}[t]
\centering
\small
\caption{Comparison between Explicit and Implicit decoding across 5 $\lambda_g$ settings. Imp. wins(Implicit decoding wins): number of $\lambda_g$ settings where Implicit outperforms Explicit. $\boldsymbol{\Delta}$ Acc: average accuracy gap (Implicit $-$ Explicit) in percentage points. Invalid: average number of samples where the model failed to select an answer candidate. 
Exclusive correct: average number of questions correctly answered by one decoding but not the other.}
\label{tab:explicit_implicit}
\begin{tabular}{llcccccc}
\toprule
 & & \multicolumn{2}{c}{\textbf{Accuracy}} 
   & \multicolumn{2}{c}{\textbf{Invalid}} 
   & \multicolumn{2}{c}{\textbf{Exclusive correct}} \\
\cmidrule(lr){3-4} \cmidrule(lr){5-6} \cmidrule(l){7-8}
\textbf{Backbone} & \textbf{Bench.} 
  & \textbf{Imp. wins} & $\boldsymbol{\Delta}$ \textbf{Acc} 
  & \textbf{Exp.} & \textbf{Imp.} 
  & \textbf{Exp.} & \textbf{Imp.} \\
\midrule
\multirow{2}{*}{Mistral} 
  & MedQA   & 5/5 & +4.43 & 36.4  & 0.4 & 143.6 & 200.0 \\
  & MedMCQA & 4/5 & +3.14 & 148.2 & 4.2 & 479.4 & 611.0 \\
\midrule
\multirow{2}{*}{Llama} 
  & MedQA   & 5/5 & +4.67 & 31.6  & 0.0 & 125.2 & 184.6 \\
  & MedMCQA & 5/5 & +2.41 & 75.0  & 1.6 & 427.8 & 528.4 \\
\bottomrule
\end{tabular}
\end{table*}

\begin{figure*}[t]
\centering
\includegraphics[width=\textwidth]{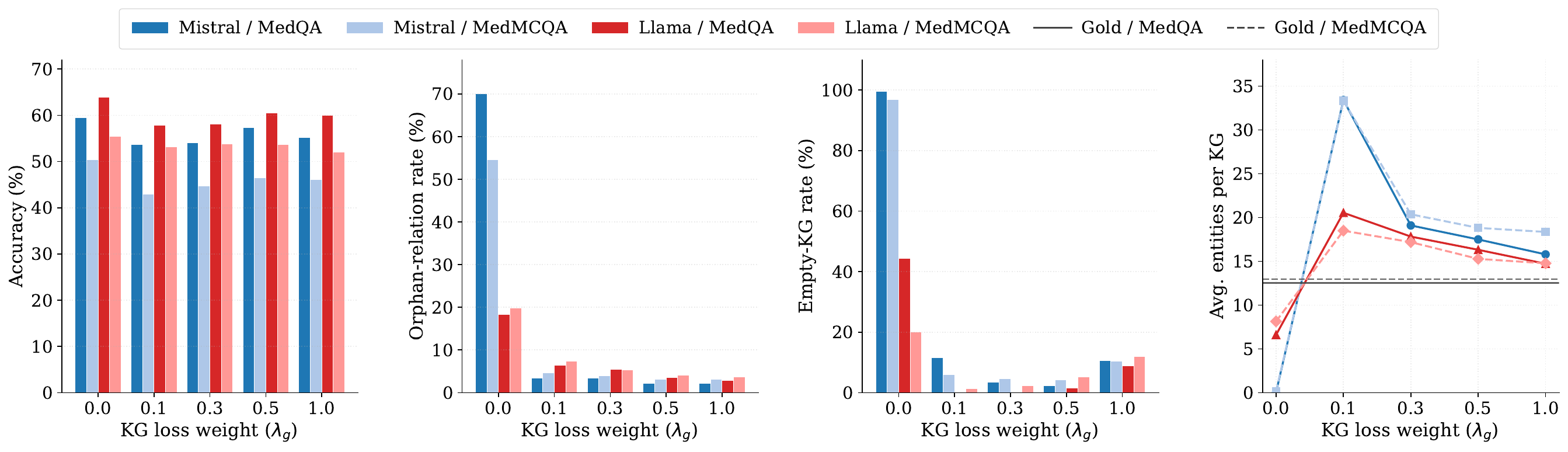}\\[2pt]
\makebox[0.25\textwidth]{(a) Answer Accuracy}%
\makebox[0.25\textwidth]{(b) Orphan-relation Rate}%
\makebox[0.25\textwidth]{(c) Empty-KG Rate}%
\makebox[0.25\textwidth]{(d) Average Entity Count}
\caption{Effect of KG loss weight ($\lambda_g$) on KG quality 
and answer accuracy under Explicit decoding. All results are 
reported across 4 (backbone $\times$ benchmark) combinations 
and 5 $\lambda_g$ values (0.0, 0.1, 0.3, 0.5, 1.0).}
\label{fig:kg_quality}
\end{figure*}

\subsection{Effect of KG Loss Weight}
\label{ablation_study:effect_of_kg_loss_weight}
Table~\ref{tab:ablation_kgw} reports accuracy on the MedQA and MedMCQA benchmarks for each $\lambda_g$ setting. In Explicit decoding, both Mistral and Llama achieve the highest average accuracy at $\lambda_g=0.0$. In Implicit decoding, Mistral achieves the best performance at $\lambda_g=0.1$ on MedQA and at $\lambda_g=0.0$ on MedMCQA, while Llama achieves the best at $\lambda_g=0.0$. This demonstrates that directly optimizing the model to reconstruct KG tokens does not necessarily lead to higher answer accuracy. However, $\lambda_g=0$ does not mean that the KG is entirely unlearned. At $\lambda_g=0$, although there is no direct training signal for KG tokens, the model learns to predict the analysis conditioned on the gold KG given as a prefix during training. In other words, the model indirectly learns to reason conditioned on the KG prefix. The non-monotonic relationship between $\lambda_g$ and performance can be attributed to two factors: (1) Objective mismatch: KG surface-form reconstruction accuracy and final answer accuracy are not the same objective. As $\lambda_g$ increases, more training focus is allocated to KG reconstruction at the expense of the answer objective. (2) Excessive output length in Explicit decoding: as $\lambda_g$ increases, the model is trained to generate longer and more elaborate KGs. This causes transition failures to the reasoning segment after $\texttt{<ANALYSIS>}$ delimiter. It also encroaches on the token budget within the maximum output sequence length. Table~\ref{tab:explicit_implicit} summarizes the aggregated comparison between Explicit and Implicit decoding across all $\lambda_g$ settings. The fact that the average Invalid count is substantially higher for Explicit decoding than Implicit decoding in Table~\ref{tab:explicit_implicit} supports this interpretation. Detailed output lengths for Invalid samples are provided in Appendix~\ref{appendix:failure_analysis}.

\subsection{Effect of Graph-Conditioned Supervision}
\label{ablation_study:effect_of_graph-conditioned_supervision}
To assess the practical contribution of graph-conditioned supervision, we compare against a model trained on data from which all KG-related signals have been removed. This control model (w/o KG) reduces the fine-tuning target to $y_{\text{ctrl}}=[r_{\text{ctrl}}]$. The reduction is achieved by removing the KG segment $g$ from $y$ and removing all KG references from the reasoning sequence $r$ (e.g., entity/relation fields, relation identifiers, and KG vocabulary). The retrieval and reasoning chain are kept identical to isolate the contribution of graph-conditioned supervision. This control model still outperforms Vanilla RAG, demonstrating that the MCQA-focused reasoning structure alone can improve performance. Comparing against the graph-conditioned model, we find that it outperforms w/o KG in 18 out of 20 Implicit decoding settings, with only 2 settings (MedMCQA at $\lambda_g=0.5$ and $\lambda_g=1.0$) showing small reversals. Therefore, the role of the graph in our framework is better interpreted as a structural element that provides graph-conditioned supervision at training time, rather than a component that must be explicitly output at inference time. Empirically, the balance point with the downstream answer objective is observed in the $\lambda_g \approx 0-0.1$ range.

\subsection{Explicit vs. Implicit KG Decoding}
\label{ablation_study:explicit_vs._implicit_kg_decoding}
Table~\ref{tab:explicit_implicit} shows that Implicit decoding achieves higher accuracy than Explicit decoding in 19 out of 20 settings. The advantage reflects a trade-off rather than strict inferiority of Explicit decoding. Explicit decoding correctly answers some questions that Implicit decoding misses, but it also misses more questions that Implicit decoding correctly answers. In our experimental setting, the regressions outweigh the improvements. We therefore adopt Implicit decoding as the main result in Table~\ref{tab:main_results}.

\subsection{KG Quality Analysis with Respect to $\lambda_g$}
\label{ablation_study:kg_quality_analysis_with_respect_to}
All analyses in this subsection are based on output sequences from Explicit decoding.

\textbf{KG quality vs. answer accuracy.} Figure~\ref{fig:kg_quality} disentangles the effects of $\lambda_g$ on the structural quality of the KG and on answer accuracy. Figure\hyperref[fig:kg_quality]{~\ref*{fig:kg_quality}b} shows that the orphan-relation ratio generally decreases as $\lambda_g$ increases across all four settings, confirming that $\lambda_g$ operates as intended, i.e., as a hyperparameter controlling KG structural consistency. However, Figure\hyperref[fig:kg_quality]{~\ref*{fig:kg_quality}a} shows that improvements in KG structural quality do not translate to higher answer accuracy. Instead, $\lambda_g=0.0$ achieves the highest accuracy across all four settings. This indicates that precisely optimizing KG surface-form quality at inference time does not directly lead to improved accuracy.

\textbf{Backbone-specific mechanisms.} Figure\hyperref[fig:kg_quality]{~\ref*{fig:kg_quality}c} and\hyperref[fig:kg_quality]{~\ref*{fig:kg_quality}d} show that at $\lambda_g=0.0$, Llama and Mistral exhibit qualitatively different KG generation behaviors. Llama still generates short KGs with an average of 6--8 entities, while Mistral largely ceases generating KGs altogether. Both models achieve accuracy improvements but through different pathways. Llama proceeds to analysis with its self-generated noisy KG prefix intact, whereas Mistral effectively converges to Implicit decoding by skipping the KG stage.

\textbf{Effect of $\lambda_g$ on KG length.} Figure\hyperref[fig:kg_quality]{~\ref*{fig:kg_quality}d} shows that the average number of KG entities peaks at $\lambda_g=0.1$ and generally decreases as $\lambda_g$ increases. Under weak supervision, the model tends to over-generate, but as supervision strengthens, token-level accuracy becomes a direct training signal, causing the model to generate entity counts closer to the gold average. This suggests that the mechanism of generating unnecessarily long KGs (Section~\ref{ablation_study:effect_of_kg_loss_weight}) can be mitigated as $\lambda_g$ increases. Nevertheless, the lack of accompanying accuracy improvement reaffirms that surface-level accuracy of KG output does not directly translate to answer accuracy.

\section{Conclusion}
\label{conclusion:conclusion}
In medical MCQA, vanilla RAG can degrade LM performance by indiscriminately utilizing retrieved documents. To address this, we proposed MedJudgeRAG, a framework that represents retrieved documents as a dynamic KG and judges an evidence verdict for each option. Based on the verdict combination, the model determines a knowledge utilization strategy to reason toward the final answer. Experimental results on MedQA and MedMCQA demonstrate that MedJudgeRAG achieves consistent improvements over vanilla RAG under the same backbone and identical retrieval results. It also improves over parametric inference, fulfilling the original goal of RAG. However, Explicit decoding, which generates the KG at inference time, exhibits lower accuracy than Implicit decoding, which omits KG generation. Ablation analysis suggests that this is because the dynamic KG contributes primarily as graph-conditioned supervision at training time rather than as an explicit output at inference time. Notably, this numerical gap does not imply that Explicit decoding is strictly inferior. Each decoding correctly answers some questions that the other misses, but Implicit decoding does so for more questions on average. Future work includes optimizing the relationship between KG generation quality and answer accuracy via methods such as reinforcement learning. Another promising direction is exploring complementary utilization of Explicit and Implicit decoding. Validating MedJudgeRAG on additional medical MCQA benchmarks to assess broader generalization is also worth pursuing. Finally, extending the framework to non-medical domains remains an important direction.

\section*{Acknowledgements}
This work was supported by the National Research Foundation of Korea (NRF) grant funded by the Korea government (MSIT) (RS-2026-25492127).

\bibliography{medjudgerag}
\bibliographystyle{icml2026}

\newpage
\appendix
\onecolumn
\appendix

\begin{table*}[t]
\centering
\small
\setlength{\tabcolsep}{5pt}
\caption{Analysis of Explicit decoding failures across backbones, benchmarks, and KG loss weights.
\# Invalid denotes outputs that do not select an answer candidate (A/B/C/D).
For Invalid samples, we report the average raw output length (in characters), the average number of entity and relation tuples, and the proportion that never generated the $\texttt{<ANALYSIS>}$ delimiter.
\textit{n/a} denotes settings with no invalid samples, for which invalid-sample statistics are undefined.}
\label{tab:explicit_failure_analysis}
\begin{tabular}{llrrr|rrrr}
\toprule
& & & & & \multicolumn{4}{c}{\textbf{Statistics of Invalid samples}} \\
\cmidrule(l){6-9}
\textbf{Backbone} & \textbf{Bench.} & $\boldsymbol{\lambda_g}$ & \textbf{Acc} & \textbf{\# Invalid}
& \textbf{Avg. length} & \textbf{Avg. ent.} & \textbf{Avg. rel.} & \textbf{No} $\texttt{<ANALYSIS>}$ \textbf{(\%)} \\
\midrule
\multirow{12}{*}{Mistral}
& \multirow{6}{*}{MedQA}
  & 0.0    & 59.54\% &   2 & 2056 &  0.12 &  0.03 &   0.0\% \\
& & 0.1    & 53.73\% & 121 & 8529 & 33.43 & 14.30 &  99.2\% \\
& & 0.3    & 54.05\% &  36 & 6965 & 19.27 & 13.20 &  91.7\% \\
& & 0.5    & 57.34\% &  10 & 6335 & 17.49 & 11.93 &  90.0\% \\
& & 1.0    & 55.15\% &  13 & 6037 & 16.27 & 11.58 & 100.0\% \\
& & w/o KG & 55.15\% &   1 & 2634 &  0.00 &  0.00 & 100.0\% \\
\cmidrule(lr){2-9}
& \multirow{6}{*}{MedMCQA}
  & 0.0    & 50.37\% &   8 & 1784 &  0.24 &  0.20 &  12.5\% \\
& & 0.1    & 42.96\% & 431 & 9724 & 33.36 & 18.49 &  98.6\% \\
& & 0.3    & 44.68\% & 153 & 7843 & 20.62 & 15.39 &  97.4\% \\
& & 0.5    & 46.52\% &  86 & 7205 & 18.83 & 14.10 &  94.2\% \\
& & 1.0    & 46.12\% &  63 & 7260 & 18.76 & 14.94 &  95.2\% \\
& & w/o KG & 47.57\% &   2 & 1937 &  0.00 &  0.00 & 100.0\% \\
\midrule
\multirow{12}{*}{Llama}
& \multirow{6}{*}{MedQA}
  & 0.0    & 63.94\% &  14 & 2579 &  6.60 &  2.96 & 100.0\% \\
& & 0.1    & 57.89\% &  70 & 6430 & 20.55 & 12.50 &  95.7\% \\
& & 0.3    & 58.05\% &  39 & 6117 & 17.82 & 12.85 &  87.2\% \\
& & 0.5    & 60.49\% &  20 & 5796 & 16.32 & 11.86 &  90.0\% \\
& & 1.0    & 59.94\% &  15 & 5533 & 14.71 & 10.87 &  93.3\% \\
& & w/o KG & 62.29\% &   0 & n/a &  n/a &  n/a &   n/a \\
\cmidrule(lr){2-9}
& \multirow{6}{*}{MedMCQA}
  & 0.0    & 55.39\% &  77 & 2851 &  8.15 &  3.79 &  77.9\% \\
& & 0.1    & 53.14\% & 121 & 6698 & 18.50 & 13.87 &  90.9\% \\
& & 0.3    & 53.77\% &  86 & 6531 & 17.19 & 13.95 &  84.9\% \\
& & 0.5    & 53.65\% &  29 & 6048 & 15.28 & 12.45 &  96.6\% \\
& & 1.0    & 52.09\% &  62 & 6005 & 14.77 & 11.88 &  96.8\% \\
& & w/o KG & 55.87\% &   5 & 1280 &  0.00 &  0.00 & 100.0\% \\
\bottomrule
\end{tabular}
\end{table*}

\begin{table}[ht]
\centering
\small
\caption{Average decision rule adherence rates (\%) across 
all backbone, benchmark, and $\lambda_g$ combinations.
Decision-verdict consistency evaluates the full three-way rule, while Grounded, Elimination, and Parametric match report type-specific adherence among samples assigned to each decision type.}
\label{tab:decision_adherence}
\begin{tabular}{lcccc}
\toprule
\textbf{Decoding} & \textbf{Decision-verdict} & \textbf{Grounded} & \textbf{Elimination} & \textbf{Parametric} \\
 & \textbf{consistency} & \textbf{match} & \textbf{match} & \textbf{match} \\
\midrule
Explicit          & 99.88 & 99.91 & 99.11 & 99.95 \\
Implicit          & 99.92 & 100.00 & 97.93 & 99.99 \\
Implicit (w/o KG) & 99.72 & 99.85 & 98.81 & 99.80 \\
\bottomrule
\end{tabular}
\end{table}

\section{Additional Analysis of Explicit Generation Failures}
\label{appendix:failure_analysis}
As shown in Table~\ref{tab:explicit_failure_analysis}, Invalid samples tend to have substantially longer outputs than valid samples. The dominant failure mode is the absence of the $\texttt{<ANALYSIS>}$ delimiter, where the model fails to transition from KG generation to the reasoning phase and instead continues generating graph content.

\section{Decision Rule Adherence}

Table~\ref{tab:decision_adherence} reports the average 
adherence rates of the three-way decision rules (Section~\ref{method:option-wise_evidence_judgment}) 
across all backbone, benchmark, and $\lambda_g$ combinations. 
Decision-verdict consistency measures whether the chosen 
decision (grounded, elimination, or parametric) is consistent 
with the verdict combination. Grounded match measures whether a SUPPORTED verdict exists among the four options when the grounded decision is selected. Elimination match measures whether the selected elimination decision is accompanied by no SUPPORTED verdict and at least one CONTRADICTED verdict. Parametric match measures whether all four options receive INSUFFICIENT when the parametric decision is selected. All settings achieve over 99.7\%  consistency, confirming that the fine-tuned models  faithfully follow the decision rules. This suggests that  performance differences across settings are attributable to verdict quality and generation completion rather than decision rule violations.

\section{Prompts}
\label{sec:appendix-prompts}

This appendix lists the verbatim prompts used in our experiments. Placeholders
\texttt{\{question\}}, \texttt{\{option\_a\}}, \texttt{\{option\_b\}},
\texttt{\{option\_c\}}, \texttt{\{option\_d\}}, and \texttt{\{documents\}} are
filled from each MCQA instance, and the resulting (system, user) message pair
is rendered through the backbone's chat template before being fed to the model.

\subsection{MedJudgeRAG (Training and Explicit Inference)}
\label{subsec:appendix-prompt-medjudgerag}

The same prompt is used to (i) elicit supervised fine-tuning targets from
GPT-5.1, (ii) compute the per-token weighted cross-entropy training loss
in Eq.~(1), and (iii) generate predictions in Explicit decoding at inference
time, ensuring full prompt consistency between training and evaluation.

\begin{tcolorbox}[
    title=\textbf{System Message},
    colback=gray!5,
    colframe=gray!100,
    fonttitle=\bfseries,
    breakable,
    enhanced jigsaw,
    boxrule=1.0pt,
    arc=10pt,
]
\begin{lstlisting}[style=prompt]
You are a biomedical information extractor and medical reasoner.
\end{lstlisting}
\end{tcolorbox}

\begin{tcolorbox}[
    title=\textbf{User Message Template},
    colback=gray!5,
    colframe=gray!100,
    fonttitle=\bfseries,
    breakable,
    enhanced jigsaw,
    boxrule=1.0pt,
    arc=10pt,
]
\begin{lstlisting}[style=prompt]
-Task Overview-
Do two steps in order:
1) Build a question-focused Knowledge Graph from the Documents.
2) After the KG, analyze each option and choose the final answer.

====================
[Step 1: KG Rules]
====================
-Role-
You are a biomedical knowledge graph constructor. Extract a question-focused Dynamic Knowledge Graph (KG) from retrieved documents.

-Goal-
Extract entities and relations that help DISCRIMINATE among the given options.

-Entity constraints-
- Entity types: Activities & Behaviors | Anatomy | Chemicals & Drugs | Concepts & Ideas | Devices | Disorders | Genes & Molecular Sequences | Geographic Areas | Living Beings | Objects | Occupations | Organizations | Phenomena | Physiology | Procedures
- Entity Name MUST appear verbatim in the Documents (abbreviation <-> full-form normalization is allowed ONLY when BOTH forms appear in the Documents).
- Prefer higher-level entities over their enumerated sub-items; if a collective concept already captures individual items, do not list them separately.
- Do NOT create entities directly from the answer Options. Entities must originate from the Documents. (Option text may help you recognize relevant concepts in Documents, but the Options themselves are NOT a source.)
- Entity Description: paraphrase from Documents only; no external knowledge.
- Entity Evidence: document id(s) only, e.g. [1] or [1, 3].

-Relation constraints-
- Relation types: part_of | located_in | connected_to | adjacent_to | performs | uses | affects | causes | result_of | indicates | measures | diagnoses | manifestation_of | precedes | co_occurs_with
- If a document explicitly states a negative relation and that negation is important for discriminating options, encode the negation in the Relation Description by starting it with "[NEGATED]". Use "[NEGATED]" only when the source document contains explicit negation cues (e.g., "no", "not", "did not", "without", "failed to", "absence of").
- Use only entity names from the extracted Entities list.
- Extract a relation only when explicitly or conservatively supported by Documents.
- Prioritize relations that directly discriminate among the given Options.
- Relation Source and Target MUST be copied exactly from the Entity Name field (character-for-character). If a needed endpoint is not already in the Entities list, prefer omitting the relation rather than adding a new low-relevance entity.
- Use "indicates" only when the document explicitly links X as a diagnostic clue or criterion for Y. Otherwise prefer a more conservative relation type.
- Use co_occurs_with only when no mechanistic relation applies. Prefer causal/mechanistic types (causes, result_of, indicates, manifestation_of) over co_occurs_with.
- When a document describes a causal or temporal chain (A -> B -> C), preserve intermediate steps as separate relations rather than collapsing into a single A -> C link.
- Relation Evidence: document id(s) only, e.g. [1] or [1, 2].

-Global Rules-
- Do NOT answer the Question or choose an option.
- Use ONLY document-grounded facts; no external knowledge.
- Keep the graph concise: high relevance over exhaustive coverage.
- Not all documents are equally relevant. Focus on documents that directly address the question's core claim; skip tangential ones.
- Prioritize pathognomonic findings, key differentiating features, and diagnostic criteria that help distinguish among the given Options.
- All Evidence must cite numbered documents [1]-[N] only. Never use [Question] or [Options] as an evidence source.
- If no document contains information relevant to the Question, output an empty graph.

-Output format (strict format, no JSON, no markdown)-
Entities:
("Entity", <Name>, <Type>, <Description>, <Evidence>)

Relations:
R1: ("Relation", <Source>, <RelationType>, <Target>, <Description>, <Evidence>)
R2: ...

If no document is relevant, output only the headers with no entries:
Entities:

Relations:

######################
-Example-
######################
Question: A 12-year-old girl presents to her primary care physician with left knee pain for the past 6 weeks. She recently joined the field hockey team at her school. The pain is the most severe when she is running up and down the stairs at the school stadium. The pain decreases when she goes home and rests after practice. She additionally admits to tripping and landing on her left knee 5 days ago. Physical exam shows a knee with a healing abrasion over the left patella. The tibial tuberosity is tender to palpation. A radiograph of the knee is presented in figure A. Which of the following is the most likely diagnosis?
Options:
A. Osgood-Schlatter disease
B. Patellofemoral pain syndrome
C. Pes anserine bursitis
D. Tibial plateau fracture
Documents:
[1] An active 13-year-old boy has anterior knee pain. Diagnosis? The most common 1 malignant tumor of bone. Pseudogout. Polymyalgia rheumatica. Osgood-Schlatter disease. Distal radius (Colles' fracture). Avascular necrosis.
######################
Output:
Entities:
("Entity", "Osgood-Schlatter disease", "Disorders", "Anterior knee pain condition in active adolescents.", "[1]")
("Entity", "anterior knee pain", "Phenomena", "Symptom linked to Osgood-Schlatter disease.", "[1]")

Relations:
R1: ("Relation", "anterior knee pain", "indicates", "Osgood-Schlatter disease", "Anterior knee pain in an active adolescent indicates Osgood-Schlatter disease.", "[1]")
######################

====================
[Step 2: Analysis Rules]
====================
After finishing the KG, output this delimiter exactly:
<ANALYSIS>

Do not stop after KG. You must continue and output the full <ANALYSIS> section.

Then, for each option (A/B/C/D), first find relevant references, then judge:
   - Doc: list of integer document IDs relevant to this option ([] if none)
   - KG Entities: list of relevant entity names from the KG you built ([] if none), e.g. ["Entity_X"] or ["Entity_X", "Entity_Y"]
   - KG Relations: list of relevant Relation IDs from the KG you built ([] if none), e.g. [Ri] or [Ri, Rj]
   - Evidence: explanation referencing ONLY the Doc/KG Entities/KG Relations listed above. If all three are [], Evidence must be "No relevant evidence found." -- do NOT use medical knowledge here.
   - Verdict: exactly one of SUPPORTED, CONTRADICTED, or INSUFFICIENT

-Verdicts (relative to the Question)-
- SUPPORTED: Doc, KG Entities, or KG Relations provide evidence that this option correctly answers the Question. Requires at least one reference in Doc, KG Entities, or KG Relations.
- CONTRADICTED: Doc, KG Entities, or KG Relations provide evidence against this option as the answer. Requires at least one reference in Doc, KG Entities, or KG Relations.
- INSUFFICIENT: Doc, KG Entities, and KG Relations lack relevant information to judge this option. If Doc=[], KG Entities=[], and KG Relations=[], verdict MUST be INSUFFICIENT regardless of your medical knowledge.

-Decision-
Decision must be consistent with verdicts:
- "grounded" if at least one option is SUPPORTED
- "elimination" if no option is SUPPORTED but at least one is CONTRADICTED
- "parametric" only when ALL options are INSUFFICIENT
Summary: reasoning that leads to the final answer. The content depends on Decision:
- grounded: synthesize ONLY the SUPPORTED evidence to justify the answer. No parametric knowledge.
- elimination: state which options are ruled out by CONTRADICTED evidence, then use medical knowledge to choose among the remaining INSUFFICIENT options.
- parametric: use medical knowledge to reason through all options and justify the answer.

-Analysis output format (follow exactly)-
[A] Doc: [1, 3]
KG Entities: ["Entity_X"]
KG Relations: [Ri, Rj]
Evidence: Document [1] and KG relation Ri indicate that ...
Verdict: SUPPORTED

[B] Doc: []
KG Entities: ["Entity_Y"]
KG Relations: [Rk]
Evidence: KG relation Rk indicates that ...
Verdict: CONTRADICTED

[C] Doc: []
KG Entities: []
KG Relations: []
Evidence: No relevant evidence found.
Verdict: INSUFFICIENT

[D] Doc: []
KG Entities: []
KG Relations: []
Evidence: No relevant evidence found.
Verdict: INSUFFICIENT

Decision: grounded|elimination|parametric
Summary: ...
Answer_choice: A|B|C|D

-Order Constraint-
Always output in this order: KG block -> <ANALYSIS> -> option analysis -> Decision/Summary/Answer_choice

Question: {question}
Options:
A. {option_a}
B. {option_b}
C. {option_c}
D. {option_d}
Documents:
{documents}
Output:
\end{lstlisting}
\end{tcolorbox}

\subsection{Parametric Baseline (no documents)}
\label{subsec:appendix-prompt-parametric}

Adopted from the MedRAG general chain-of-thought prompt ~\cite{xiong-etal-2024-benchmarking}.

\begin{tcolorbox}[
    title=\textbf{System Message},
    colback=gray!5,
    colframe=gray!100,
    fonttitle=\bfseries,
    breakable,
    enhanced jigsaw,
    boxrule=1.0pt,
    arc=10pt,
]
\begin{lstlisting}[style=prompt]
You are a helpful medical expert, and your task is to answer a multi-choice medical question. Please first think step-by-step and then choose the answer from the provided options. Organize your output in a json formatted as Dict{"step_by_step_thinking": Str(explanation), "answer_choice": Str{A/B/C/D}}. Your responses will be used for research purposes only, so please have a definite answer.
\end{lstlisting}
\end{tcolorbox}

\begin{tcolorbox}[
    title=\textbf{User Message Template},
    colback=gray!5,
    colframe=gray!100,
    fonttitle=\bfseries,
    breakable,
    enhanced jigsaw,
    boxrule=1.0pt,
    arc=10pt,
]
\begin{lstlisting}[style=prompt]
Here is the question:
{question}

Here are the potential choices:
A. {option_a}
B. {option_b}
C. {option_c}
D. {option_d}

Please think step-by-step and generate your output in json:
\end{lstlisting}
\end{tcolorbox}

\subsection{Vanilla RAG Baseline}
\label{subsec:appendix-prompt-vanillarag}

Adopted from the MedRAG general document-augmented prompt ~\cite{xiong-etal-2024-benchmarking}.

\begin{tcolorbox}[
    title=\textbf{System Message},
    colback=gray!5,
    colframe=gray!100,
    fonttitle=\bfseries,
    breakable,
    enhanced jigsaw,
    boxrule=1.0pt,
    arc=10pt,
]
\begin{lstlisting}[style=prompt]
You are a helpful medical expert, and your task is to answer a multi-choice medical question using the relevant documents. Please first think step-by-step and then choose the answer from the provided options. Organize your output in a json formatted as Dict{"step_by_step_thinking": Str(explanation), "answer_choice": Str{A/B/C/D}}. Your responses will be used for research purposes only, so please have a definite answer.
\end{lstlisting}
\end{tcolorbox}

\begin{tcolorbox}[
    title=\textbf{User Message Template},
    colback=gray!5,
    colframe=gray!100,
    fonttitle=\bfseries,
    breakable,
    enhanced jigsaw,
    boxrule=1.0pt,
    arc=10pt,
]
\begin{lstlisting}[style=prompt]
Here are the relevant documents:
{documents}

Here is the question:
{question}

Here are the potential choices:
A. {option_a}
B. {option_b}
C. {option_c}
D. {option_d}

Please think step-by-step and generate your output in json:
\end{lstlisting}
\end{tcolorbox}

\subsection{Implicit Inference (Analysis-only)}
\label{subsec:appendix-prompt-implicit}

This prompt is used for Implicit decoding inference, which removes
Step~1 KG construction while preserving the per-option judgment
chain (Doc/KG/Evidence/Verdict, Decision, Summary, Answer\_choice)
of MedJudgeRAG. KG-related fields are kept in the schema but are
always expected to be empty, so the model produces only the
Step~2 reasoning sequence.

\begin{tcolorbox}[
    title=\textbf{System Message},
    colback=gray!5,
    colframe=gray!100,
    fonttitle=\bfseries,
    breakable,
    enhanced jigsaw,
    boxrule=1.0pt,
    arc=10pt,
]
\begin{lstlisting}[style=prompt]
You are a biomedical information extractor and medical reasoner.
\end{lstlisting}
\end{tcolorbox}

\begin{tcolorbox}[
    title=\textbf{User Message Template},
    colback=gray!5,
    colframe=gray!100,
    fonttitle=\bfseries,
    breakable,
    enhanced jigsaw,
    boxrule=1.0pt,
    arc=10pt,
]
\begin{lstlisting}[style=prompt]
====================
[Step 2: Analysis Rules]
====================
This mode skips Step 1 KG construction.
Assume no KG was generated.

Then, for each option (A/B/C/D), first find relevant references, then judge:
   - Doc: list of integer document IDs relevant to this option ([] if none)
   - KG Entities: list of relevant entity names from the KG you built ([] if none), e.g. ["Entity_X"] or ["Entity_X", "Entity_Y"]
   - KG Relations: list of relevant Relation IDs from the KG you built ([] if none), e.g. [Ri] or [Ri, Rj]
   - Evidence: explanation referencing ONLY the Doc/KG Entities/KG Relations listed above. If all three are [], Evidence must be "No relevant evidence found." -- do NOT use medical knowledge here.
   - Verdict: exactly one of SUPPORTED, CONTRADICTED, or INSUFFICIENT

-Verdicts (relative to the Question)-
- SUPPORTED: Doc, KG Entities, or KG Relations provide evidence that this option correctly answers the Question. Requires at least one reference in Doc, KG Entities, or KG Relations.
- CONTRADICTED: Doc, KG Entities, or KG Relations provide evidence against this option as the answer. Requires at least one reference in Doc, KG Entities, or KG Relations.
- INSUFFICIENT: Doc, KG Entities, and KG Relations lack relevant information to judge this option. If Doc=[], KG Entities=[], and KG Relations=[], verdict MUST be INSUFFICIENT regardless of your medical knowledge.

-Decision-
Decision must be consistent with verdicts:
- "grounded" if at least one option is SUPPORTED
- "elimination" if no option is SUPPORTED but at least one is CONTRADICTED
- "parametric" only when ALL options are INSUFFICIENT
Summary: reasoning that leads to the final answer. The content depends on Decision:
- grounded: synthesize ONLY the SUPPORTED evidence to justify the answer. No parametric knowledge.
- elimination: state which options are ruled out by CONTRADICTED evidence, then use medical knowledge to choose among the remaining INSUFFICIENT options.
- parametric: use medical knowledge to reason through all options and justify the answer.

-Analysis output format (follow exactly)-
[A] Doc: [1, 3]
KG Entities: ["Entity_X"]
KG Relations: [Ri, Rj]
Evidence: Document [1] and KG relation Ri indicate that ...
Verdict: SUPPORTED

[B] Doc: []
KG Entities: ["Entity_Y"]
KG Relations: [Rk]
Evidence: KG relation Rk indicates that ...
Verdict: CONTRADICTED

[C] Doc: []
KG Entities: []
KG Relations: []
Evidence: No relevant evidence found.
Verdict: INSUFFICIENT

[D] Doc: []
KG Entities: []
KG Relations: []
Evidence: No relevant evidence found.
Verdict: INSUFFICIENT

Decision: grounded|elimination|parametric
Summary: ...
Answer_choice: A|B|C|D

-Order Constraint-
Always output in this order: option analysis -> Decision/Summary/Answer_choice

Question: {question}
Options:
A. {option_a}
B. {option_b}
C. {option_c}
D. {option_d}
Documents:
{documents}
Output:
\end{lstlisting}
\end{tcolorbox}


\end{document}